# Laser Directly Written Junctionless In-plane-Gate Neuron Thin Film Transistors with AND Logic Function

Li Qiang Zhu, Guo Dong Wu, Ju Mei Zhou, Wei Dou, Hong Liang Zhang, and Qing Wan[a]

Ningbo Institute of Materials Technology and Engineering, Chinese Academy of Sciences, Ningbo 315201, People's Republic of China

Junctionless oxide-based neuron thin-film transistors with in-plane-gate structure are fabricated at room temperature with a laser scribing process. The neuron transistors are composed of a bottom ITO floating gate and multiples of two in-plane control gates. The control gates, coupling with the floating gate, control the "on" and "off" of the transistor. Effective field-effect modulation of the drain current has been realized. AND logic is demonstrated on a dual in-plane gate neuron transistor. The developed laser scribing technology is highly desirable in terms of the fabrication of high performance neuron transistors with low-cost.

---

[a] Corresponding author. E-mail address: wanqing@nimte.ac.cn

Multigate devices have been proposed as a viable solution to complementary metal-oxide-semiconductor (CMOS) scaling issues.[1] More functions could be introduced to a single device through multi-independent gate, therefore increasing the functional density for a given area. Dual-gate thin-film transistors (TFTs) have attracted a lot of attentions in chemical and biological sensing, pixel display driver and logic circuit applications.[2-6] A highly-functional MOS transistor called neuron MOS transistor has been developed, similar to the function of biological neurons by executing a weighted sum calculation of multiple input signals and then controls the "on" and "off" states of the transistor, thus realizing the neuron function.[7-9] The device has a floating gate and multiple input control gates which are capacitively coupling with the floating gate, featuring a full compatibility with a standard CMOS process. However, due to the multiple gate/dielectric deposition and precise photolithography steps, the fabricating of such neuron transistors by standard CMOS process is expensive and time-consuming. Recently, the concept of junctionless transistors has been proposed and explored [10-12]. Compared with the conventional field-effect transistors (FETs), the unique feature of such junctionless transistors is that the channel doping is the same or comparable to that of the source and drain (S/D), therefore no sour/drain junction formation steps are needed, and the carrier transport is less sensitive to the channel interface. However, fabrication of these junctionless FETs is still rather challenging. As an interesting concept for promising functional devices, in-plane gate transistors have been developed as logic devices [13], rectifier [14, 15], negative differential resistance devices [16], etc. Laser scribing process was also proposed for source/drain patterning.[17, 18] However, masks are still needed during the TFTs fabrication.

In this work, a laser scribing method without any masks and photolithography is developed to directly write the junctionless in-plane-gate neuron TFTs arrays on glass substrates at room temperature. $SiO_2$-based solid electrolyte works as the gate dielectrics.[19, 20] The channel and source/drain electrodes are realized by a thin indium-tin-oxide (ITO) layer without any intentional source/drain junction. The bottom ITO layer works as a floating gate, while the top isolated ITO works as the in-plane control gate. Effective field-effect modulation of the drain current has been realized on both the single in-plane-gate (SG) mode and the dual in-plane-gate (DG) mode. Good electrical performance have been exhibited on SG neuron TFT: low OFF current of <3 pA, high $I_{on}/I_{off}$ ratio of >$10^6$, high electron mobility of ~12.6 $cm^2$/Vs and low subthreshold swing of 0.25 V/decade. AND-logic was experimentally demonstrated on the DG neuron TFT. The developed laser scribing technology is highly desirable in terms of the low-cost fabrication process.

Fabrication of ITO junctionless in-plane gate neuron transistors was performed at room temperature, as schematically shown in Fig.1 (a) and (b). After cleaning the substrates, a 100-nm-thick Al thin film was deposited on ITO surface by sputtering. Then, a 2-μm-thick $SiO_2$-based solid electrolyte film was deposited by plasma enhanced chemical vapor deposition (PECVD) using $SiH_4$ and $O_2$ as reactive gases. Finally, a 25-nm-thick ITO films were deposited on the $SiO_2$-electrolyte based gate dielectrics by sputtering. The ratio frequency (RF) power, Ar flow rate and the chamber pressure were set to be 100 W, 14 sccm and 0.5 Pa, respectively. Then, the neuron TFTs were fabricated on the prepared stacks of ITO/$SiO_2$-electrolyte/Al/ITO/glass substrates by adopting a laser scribing process. The focused laser beam selectively etches out the deposited thin films due to a localized thermoelastic force caused by rapid thermal expansion resulting from the pulsed laser irradiation. The laser patterned ITO arrays are with the dimension of 1 mm×0.2 mm. The capacitors (C1, C2, C3, …, Cn) are effectively coupled through bottom ITO floating gate, as shown in Fig.1 (c). The source and drain are obtained on the same ITO region, while the isolated ITO region works as the in-plane control gate. When set signals on G1 (or G1 and G2), the junctionless SG (or DG) neuron TFT is obtained. The capacitance-frequency measurement of the $SiO_2$-based solid electrolyte was performed using an Impedance Analyzer. The electrical characteristics of the junctionless SG and DG TFTs were measured with a keithley 4200 SCS semiconductor parameter analyzer at room temperature in the dark.

The output characteristics of the laser patterned junctionless SG neuron TFTs are shown in Fig.2. The nominal channel length is 0.4 mm, while the channel width is 0.2 mm (as shown in the inset of Fig.2). At low $V_{ds}$, the drain current increases linearly with drain voltage, indicating that the device has a good ohmic contact. At the higher $V_{ds}$, the drain current gradually approaches a saturated value. With a gate voltage of 2 V, the saturation current is observed to be ~5 µA at a drain voltage below 2 V.

Fig.3 shows the transfer characteristics of the junctionless SG neuron TFTs at saturation mode, fixed at $V_{ds}$ of 2V. It can be found that $I_{ds}$ can be effectively modulated by $V_{gs}$. The subthreshold swing (S) is found to be ~0.25 V/dec. The drain current on/off ratio $I_{on}/I_{off}$ is determined to be ~$3\times10^6$. A threshold voltage $V_{th}$ of -0.7 V is estimated by extrapolating the linear portion of the curves relating $I_{ds}^{1/2}$ and $V_{gs}$ to $I_{ds}^{1/2}=0$. The field-effect mobility (µ) in the saturation region can be extracted from the following equation:

$$I_{ds} = (\frac{WC_i\mu}{2L})(V_{gs}-V_{th})^2 \qquad (V_{ds} > V_{gs} - V_{th}) \qquad (1)$$

where L is the channel length, W is the channel width, and $C_i$ is the unit area capacitance of the dielectrics. Capacitor of $SiO_2$-based solid electrolyte with two in-plane ITO electrodes was measured using an ITO/$SiO_2$-electrolyte/Al/ITO in-plane test structure as shown in Fig.1 (c) with G1 and S as the electrode (results not shown here). The in-plane gate capacitance of 0.25 µF/cm$^2$ is used to calculate the field-effect mobility. The field-effect electron mobility is estimated to be ~12.6 cm$^2$/Vs.

Fig.4 shows the transfer characteristics of the junctionless DG neuron TFTs at saturation region, fixed at $V_{ds}$ of 1.5 V. Both the channel width and the channel length is ~0.2 mm (as shown in the inset of Fig.4). The drain currents are controlled by two gates (G1 and G2). G1 bias sweeps from -1.5 V to 1.5 V with the fixed G2 bias set to 1.5 V, 0 V and -1.5 V, respectively. It can be found obviously from Fig.4 that the drain current can be effectively modulated by G1 bias with fixed G2 bias of 1.5 V and 0V. The drain current can be turned on or off effectively with the drain current on/off ratio $I_{on}/I_{off}$ of above $10^5$ and above $10^4$ for G2 bias of 1.5V and 0 V, respectively. The subthreshold swing (S) is found to be ~0.25 V/dec and ~0.33 V/decade for G2 bias of 1.5 V and 0 V, respectively. A threshold voltage $V_{th}$ of -0.25 V is estimated for G2 bias of 1.5 V. All the results here indicate that the laser patterned in-plane gate junctionless neuron TFTs could operate under the low voltage of 1.5 V. Interestingly, the drain current can not be turned on effectively under the G2 bias of -1.5 V. The DG TFT performances are meaningful for logic gate applications.

The logic circuit of a dual in-plane gate (DG) neuron TFT is shown in Fig.5(a). The devices are characterized by applying different fixed potentials, corresponding to two logic states, HIGH state ("1") at 1.5 V and LOW state ("0") at -1.5 V, input directly and independently to each of the two control gates, seen the logic circuit diagram for input 1 and input 2. This in-plane gate logic operates as an AND-gate as demonstrated in Fig.5 (b). The drain current monitored are also detected as output ("1": $I_{ds}$ above 0.1 µA; or "0": $I_{ds}$ below 0.01 µA). At "00", "01", or "10", the device is in the OFF state. A low OFF current of <0.01 nA is measured. For state "11", the device is ON with a HIGH current of >0.7 µA. The operation of the logic device is explained as follows. When a HIGH signal is applied to each gate of the DG TFT (input 1 and input 2), the ITO channel works in accumulate region and drain current shifts to HIGH current. When a LOW signal is applied to either gate of input 1 or input 2, the ITO channel are effectively modulated to a depletion region and the HIGH signal on another input can not accumulate the ITO channel effectively, therefore the drain current shifts to LOW current. The results indicate a high ON/OFF ratio of ~$10^4$ between the two logic states. Though there are some small noises of below 0.1 nA, the HIGH/LOW ratio output is still over $10^3$. The HIGH/LOW ratio is high enough for reliable logic operations.

In summary, a simple laser scribing process without any mask and photolithography is implemented in fabricating junctionless in-plane gate neuron TFT arrays. Such junctionless neuron TFTs feature that the channel and the source/drain electrodes are of the same indium-tin-oxide (ITO) films without any intentional source/drain junction formation steps. The bottom ITO layer works as a floating gate, while the top isolated ITO works as the in-plane control gate. The control gates, coupling with the floating gate, control the "on" and "off" of the transistor. Effective field-effect modulation of the drain current has been realized on both the single in-plane gate (SG) mode and the dual in-plane gate (DG) mode. AND logic is demonstrated on a dual in-plane gate neuron transistor with high HIGH/LOW ratios, allowing the reliable logic operation. The developed laser scribing technology is highly desirable in terms of the fabrication of high performance neuron transistors with low-costs.

**Acknowledgements**

The authors are grateful for the financial supports from the National Program on Key Basic Research Project (2012CB933004), the National Natural Science Foundation of China (11174300, 11104288).

Figure Captions

Fig.1 (a) The obtained ITO/SiO$_2$-electrolyte/Al/ITO/Glass stack. (b) Laser scribing process results in the isolation of the top ITO films as well as the SiO$_2$-electrolyte and Al layer. (c) Schematic cross-sectional view of the laser patterned junctionless in-plane gate neuron TFTs structure. The capacitors (C1, C2, C3, …, Cn) are effectively coupled through bottom ITO floating gate.

Fig.2 Output characteristics (I$_{ds}$ *vs* V$_{ds}$) of the laser patterned junctionless SG TFTs. Inset: Top view optical image of the junctionless single in-plane-gate (SG) neuron TFTs in electrical measurement on a probe station.

Fig.3 Transfer characteristics (I$_{ds}$ *vs* V$_{gs}$ and the square root of I$_{ds}$ *vs* V$_{gs}$) of the junctionless SG neuron TFTs at saturation mode, fixed at V$_{ds}$ of 2 V.

Fig.4 The transfer characteristics of the junctionless DG neuron TFTs at saturation region, fixed at V$_{ds}$ of 1.5 V. Inset: Top view optical image of the junctionless dual in-plane-gate (DG) neuron TFTs in electrical measurement on a probe station.

Fig.5 (a) Logic circuit diagram of the DG neuron TFT. (b) AND logic operation.

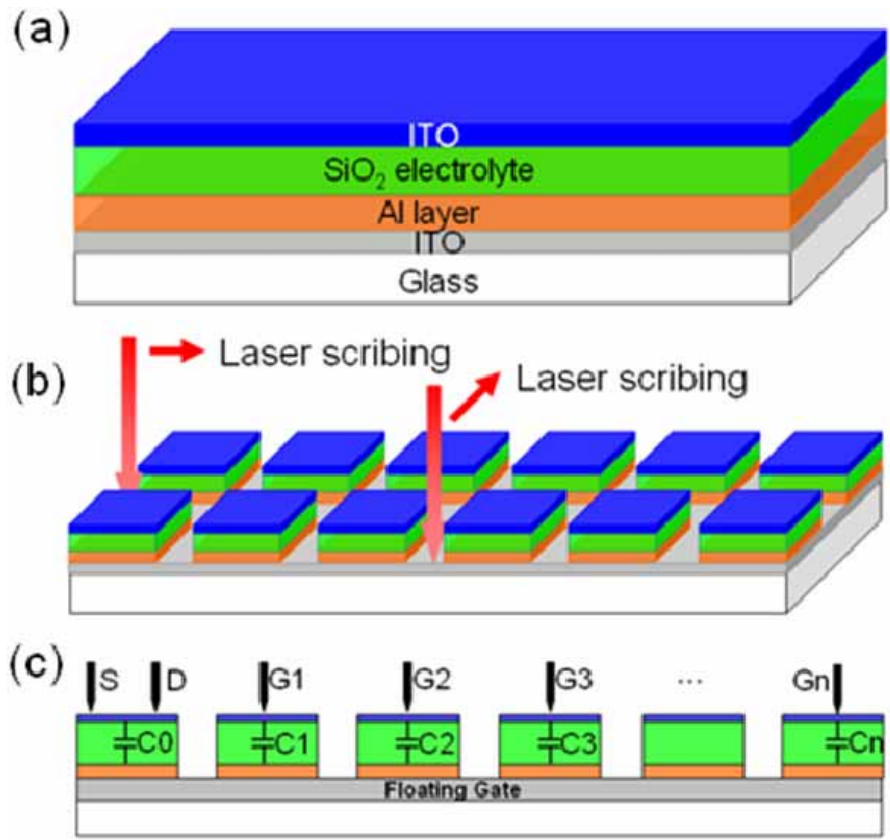

**Figure 1**

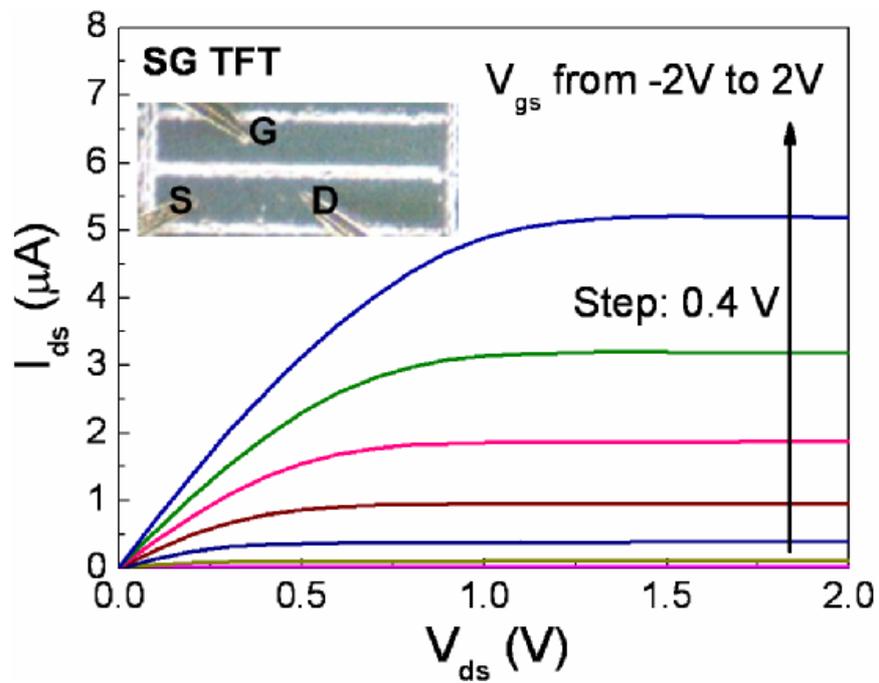

**Figure 2**

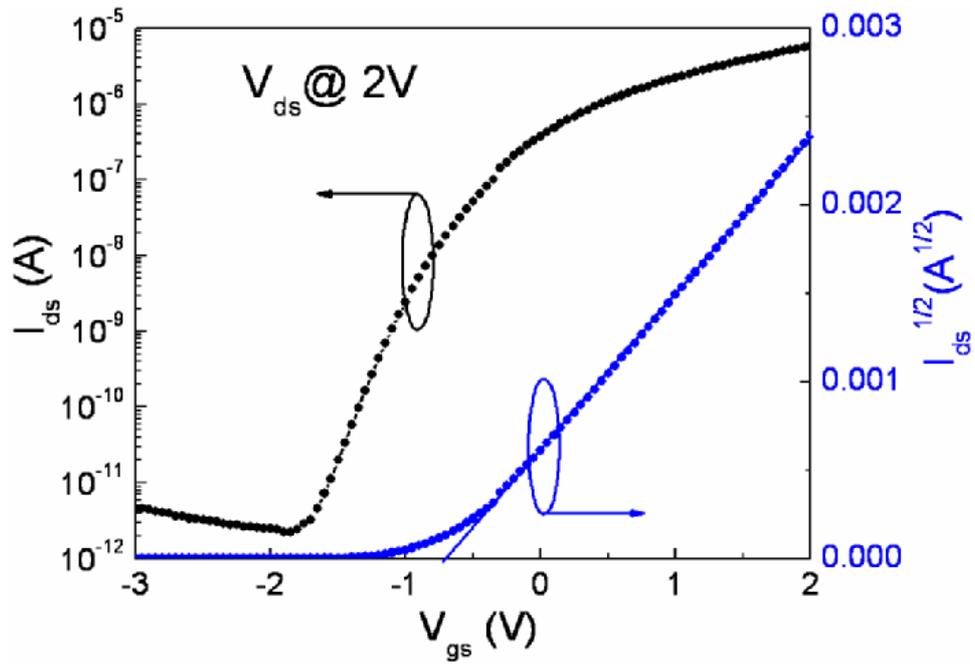

**Figure 3**

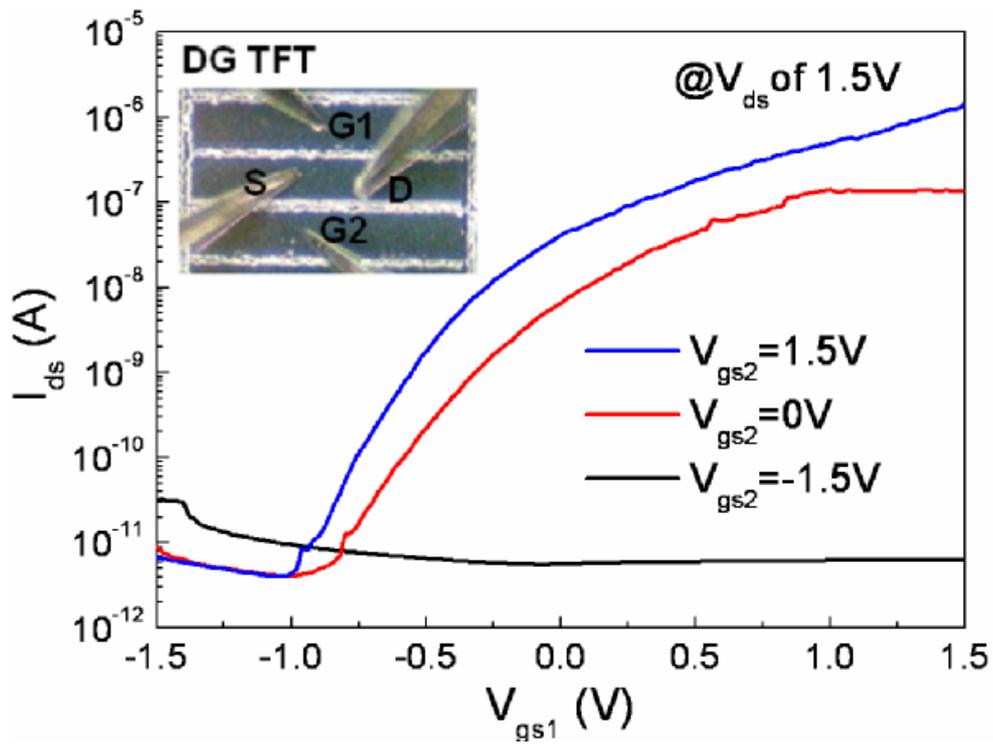

**Figure 4**

**Figure 5**